\documentclass[sn-mathphys-num]{sn-jnl}% Math and Physical Sciences Numbered Reference Style 

\usepackage{graphicx}%
\usepackage{physics}
\usepackage{multirow}%
\usepackage{amsmath,amssymb,amsfonts}%
\usepackage{amsthm}%
\usepackage{mathrsfs}%
\usepackage[title]{appendix}%
\usepackage[table]{xcolor}
\usepackage{textcomp}%
\usepackage{manyfoot}%
\usepackage{booktabs}%
\usepackage{algorithm}%
\usepackage{algorithmicx}%
\usepackage{algpseudocode}%
\usepackage{listings}%

\theoremstyle{thmstyleone}%
\theoremstyle{thmstyletwo}%
\newtheorem{example}{Example}%

\theoremstyle{thmstylethree}%

\begin{document}

\title[Article Title]{Secure and Efficient Quantum Signature Scheme Based on the Controlled Unitary Operations Encryption}


\author[1]{\fnm{Debnath} \sur{Ghosh}}\email{debnath.ghosh@research.iiit.ac.in} 

\author*[2]{\fnm{Soumit} \sur{Roy}}\email{soumit.roy@research.iiit.ac.in}

\author[1]{\fnm{Prithwi} \sur{Bagchi}}\email{prithwi.bagchi@research.iiit.ac.in} 

\author[1,2]{\fnm{Indranil} \sur{Chakrabarty}}\email{indranil.chakrabarty@iiit.ac.in} 

\author[1]{\fnm{Ashok Kumar} \sur{Das}}\email{iitkgp.akdas@gmail.com, ashok.das@iiit.ac.in} 


\affil[1]{\orgdiv{Center for Security, Theory and Algorithmic Research}, \orgname{International Institute of Information Technology}, \orgaddress{\city{Hyderabad}, \postcode{500032}, \country{India}}}

\affil*[2]{\orgdiv{Centre for Quantum Science and Technology}, \orgname{International Institute of Information Technology}, \orgaddress{\city{Hyderabad}, \postcode{500032}, \country{India}}}


\abstract{Quantum digital signatures ensure unforgeable message authenticity and integrity using quantum principles, offering unconditional security against both classical and quantum attacks. They are crucial for secure communication in high-stakes environments, ensuring trust and long-term protection in the quantum era. Nowadays, the majority of arbitrated quantum signature (AQS) protocols encrypt data qubit by qubit using the quantum one-time pad (QOTP). Despite providing robust data encryption, QOTP is not a good fit for AQS because of its susceptibility to many types of attacks. In this work, we present an efficient AQS protocol to encrypt quantum message ensembles using a distinct encryption technique, the chained controlled unitary operations. In contrast to existing protocols, our approach successfully prevents disavowal and forgery attacks. We hope this contributes to advancing future investigations into the development of AQS protocols.}


\maketitle

\section{Introduction}\label{sec1}
Quantum cryptography is one of the fastest-growing areas under the broad umbrella of quantum information theory, constantly updating itself with new theoretical protocols and security proofs. On the experimental side, we are gradually moving from proof of principle lab demonstrations to infield technological implementation  \cite{dowling2003quantum}. Quantum cryptography as a whole is no longer restricted to key generation  \cite{bennett1992quantum,ekert1991quantum} by harnessing properties like indistinguishability of quantum states \cite{nielsen2010quantum} and bell violations  \cite{bell1964much} to name a few, but also going beyond to the domains like quantum random number generator \cite{mannalatha2023comprehensive}, quantum digital signature  \cite{gottesman2001quantum}, quantum secret sharing and controlled state reconstruction  \cite{hillery1999quantum,singh2024controlled,abrol2024secret,adhikari2010probabilistic} and quantum money  \cite{wiesner1983conjugate}. However in the journey there are also some significant hurdles to tasks like bit commitment  \cite{lo1997quantum}, secure two party computation using quantum communication, quantum rewinding  \cite{lai2022quantum}, oblivious transfer \cite{santos2022quantum} and position based quantum cryptography where there are reported inadequacies from the quantum cryptography side. 

Quantum mechanical effects other than being cardinal reasons for various information processing tasks  \cite{bennett1993teleporting,horodecki1996teleportation,sohail2023teleportation,bennett1992communication}, is key to the private nature of the quantum information itself. This leads to a new era of security and privacy, fundamentally different from existing systems. However, on the flip side, it stimulates quantum attacks on the existing secure systems through its access to quantum computers  \cite{shor1999polynomial}. As a consequence of this, a significant effort was made in creating classical protocols that will be secure from potential quantum attacks \cite{bernstein2017post}. The idea of quantum cryptography was further extended to the model of device-independent cryptography, tolerating a realistic amount of noise. Several researches were carried out in this domain  \cite{nadlinger2022experimental,zhang2022device}.

A digital signature is a cryptographic method that can be utilized to authenticate the source of the transmitted data and verify the integrity of the data. They are widely used in email security, software distribution, financial transactions, legal document signing, healthcare records protection, and government services. Over the past decades, researchers have proposed numerous classical public-key digital signature protocols. Typically, these protocols' security relies on computational hardness assumptions, like the discrete logarithm problem and the prime factorization problem. As quantum computing advances, classical digital signature schemes face security risks due to their vulnerability to quantum algorithms like Shor's, which can efficiently break RSA and DSA-based systems. Quantum digital signature schemes offer unconditional security based on quantum mechanics, ensuring authenticity and integrity against quantum adversaries. Additionally, the development of post-quantum cryptographic algorithms, such as lattice-based or hash-based signatures, aims to secure communications against both classical and quantum attacks. This transition is essential to maintain robust security in the face of emerging quantum threats.

In recent years, the advancement of quantum signature schemes has significantly elevated the importance of the quantum cryptography field. In 2001, Gottesman and Chuang \cite{gottesman2001quantum} developed a technique for digitally signing and verifying messages, utilizing a quantum one-way function alongside the quantum swap test. Since that time, numerous efforts have been made to broaden the range of signable messages to include both known and unknown quantum messages. Many of these are arbitrated quantum signature (AQS) schemes, where disputes between the signers and verifiers are resolved by an arbitrator, who is presumed to be a reliable intermediary party. In 2002, Zeng and Keitel \cite{ZengKeitel} introduced an arbitrated quantum signature (AQS) scheme that relies on the correlation of the GHZ state combined with a quantum symmetric encryption technique. Since then, numerous quantum signature schemes \cite{Zhang2013,Yang2016,Su2013,Wang2016,XIN201923} have been proposed in the literature. In 2013, Zhang \textit{et al.} \cite{Zhang2013} analyzes the security flaws in existing arbitrated quantum signature (AQS) protocols, particularly regarding forgery attacks by the receiver. It proposes enhanced encryption methods to improve security against such attacks while maintaining the advantages of AQS. In 2016, Yang \textit{et al.} \cite{Yang2016} introduced an efficient arbitrated quantum signature scheme employing the cluster states, ensuring security and authenticity in quantum communication. In 2020, Xin \textit{et al.} \cite{XIN202016} introduced an ``identity-based quantum signature protocol'' utilizing the Bell states. It eliminates the need for quantum swap tests and long-term quantum memory during verification, enhancing efficiency and practicality. Later in the year 2020, Xin \textit{et al.} \cite{Xin2020} suggested a public key quantum signature technique that is identity-based and uses the Hadamard operator and Bell states. Both these schemes address the issue of forged signatures by the Private Key Generator (PKG) and resolve disputes arising from the loss of quantum signatures. Over the years, multiple arbitrated quantum signature protocols have been proposed with improved security enhancements.

In the landscape of post-quantum cryptography, several signature schemes \cite{Ducas2017,fouque2018falcon,Beullens2022,Marius2024} have emerged as frontrunners, most notably the NIST-standardized CRYSTALS-Dilithium, Falcon, and SPHINCS+. These schemes rely on different hardness assumptions, including module-lattices and hash-based structures, and offer varying trade-offs between security, performance, and signature size. More recent efforts like HAWK and GPV-Sign have pushed the frontier of efficiency and compactness in lattice-based schemes. Alongside these classical efforts, more recent advances in arbitrated quantum signature (AQS) schemes \cite{Sunil2024,mohanty2024} have introduced stronger security models and experimental validation. These advancements highlight the critical need for both classical and quantum-native signature schemes to evolve in response to quantum threats.

\subsection{Research motivation and Contributions}
Arbitrated quantum signature (AQS) protocols have historically been vulnerable to various attacks  \cite{Choi_2011,gao2011,sun2011}, particularly those exploiting the structure of quantum one-time pad encryption. For instance, Gao \textit{et al.}  \cite{gao2011} observed that the quantum one-time pad's (QOTP) encryption features and the four Pauli operators' commutative nature enable the recipient, Bob, to counterfeit the signer's signature during known message attacks, exposing a fundamental limitation of quantum one-time pad (QOTP)-based AQS schemes. Choi \textit{et al.}  \cite{Choi_2011} and Zhang \textit{et al.}  \cite{Zhang2013Re} further identified weaknesses in early AQS protocols, leading to improvements such as the chained CNOT-based encryption proposed by Li and Shi  \cite{Li2015CNOT}. Subsequent enhancements by Zhang \textit{et al.}  \cite{Zhangimproved} improved resistance to forgery but remained limited in terms of leveraging quantum properties such as phase manipulation or superposition for increased security.

Zhang \textit{et al.}'s protocol \cite{Zhangimproved}, while effective in reducing vulnerability to Pauli attacks through key-controlled CNOT chains and Hadamard gates, still relies heavily on operations that act only on the bit-flip structure. CNOT operations are inherently linear and Clifford-limited, which introduces structural regularities that attackers could potentially exploit. In contrast, the controlled-unitary operations used in our proposed scheme can implement non-Clifford rotations based on secret keys, enabling nonlinear and non-commutative encryption paths. Our protocol thus exploits more general controlled-unitary gates to manipulate both amplitude and phase components of the quantum message. This provides a higher degree of encryption randomness and non-commutativity compared to Zhang \textit{et al.}'s \cite{Zhangimproved} CNOT-based structure, helping to resist not only Pauli-based attacks but also more general quantum forgeries. By eliminating structural symmetry that may otherwise be exploitable, we enhance both theoretical and practical robustness.

In this work, our goal is to address these shortcomings by designing a more secure AQS protocol that is resistant to receiver-side forgery attacks and leverages quantum mechanical features more effectively. The main contributions of our paper are as follows:
\begin{itemize}
    \item We propose a novel quantum encryption algorithm for AQS protocols that replaces chained CNOT operations with chained controlled-unitary operations, offering greater generality and tunability.
    \item We analyze the security properties of the proposed protocol, demonstrating its resistance to Pauli operator-based forgery by the receiver along with other attacks like impersonation attacks and man-in-the-middle attacks.
    \item We implement the proposed protocol on real quantum hardware using IBM Quantum’s back-end platform, demonstrating its practical viability under realistic noise conditions.
    \item We provide a comparative analysis of our protocol against existing AQS schemes, highlighting improvements in both security design and experimental feasibility.
\end{itemize}
This work contributes a new direction in AQS protocol development by integrating unitary gate-based encryption techniques with practical implementation, addressing both theoretical limitations and real-world applicability.

\subsection{Structure of Article}
The structure of this work is as follows: Section  \ref{sec2} describes the mathematical preliminaries, which include One-Time Pad (OTP), representation of unitary gate and key-controlled chained CU gate-based encryption method. Section  \ref{sec3} reviews related articles and highlights the research gaps within them. Section  \ref{sec4} presents the proposed quantum signature scheme. Section  \ref{sec5} provides a security analysis and an in-depth discussion of the proposed scheme. In Section  \ref{sec6}, we test our proposed protocol on a quantum simulator that emulates an IBM Quantum back-end device, followed by implementation and execution on a real back end. Section  \ref{sec7} includes a comprehensive comparison of the proposed scheme with existing related schemes. Finally, the paper is concluded in Section  \ref{sec8}.

\section{Mathematical Preliminaries}
\label{sec2}
In this section, we present the basic concepts that are relevant and necessary to comprehend our proposed scheme.

\subsection{One-Time Pad (OTP)}
In cryptography, the OTP \cite{miller1882telegraphic,FrankmillerOTP} is a symmetric encryption mechanism that was first introduced by a Californian banker named Frank Miller in 1882. He released a work titled ``Telegraphic Code to Insure Privacy and Secrecy in the Transmission of Telegrams'' \cite{miller1882telegraphic}, outlining the initial one-time pad system. In OTP, the sender obtains the ciphertext $\mathcal{C}= \mathcal{M} \oplus k$ by XORing the bits of the message $\mathcal{M}$ with the bits of the random secret key $k$, which must be as long as the message $\mathcal{M}$. The encrypted message $\mathcal{C}$ is then sent to the receiver. Upon receiving $\mathcal{C}$, the receiver computes $\mathcal{C} \oplus k$ to retrieve the original plaintext $\mathcal{M}$. If the key is truly random and never used again, the OTP will remain unbreakable. It has been proven that the OTP is unconditionally secure \cite{Shannon_unconditionally_secure}.

\subsection{Unitary gate}
The $U$ gate is the generic single-qubit rotation gate. The $U$ gate's parameterized matrix representation is provided by
\begin{eqnarray}\label{Ugate}
    U(\theta, \phi, \lambda) = \begin{pmatrix}
    \cos\frac{\theta}{2} & -e^{i\lambda}\sin\frac{\theta}{2} \cr
    e^{i\phi}\sin\frac{\theta}{2} & e^{i(\phi+\lambda)}\cos\frac{\theta}{2} 
    \end{pmatrix},
\end{eqnarray}
where the three Euler angles are $\phi, \theta$, and $\lambda$.

\subsection{Key-controlled chained CU gate-based encryption}
The ``Quantum One-Time Pad (QOTP) encryption technique'' that is used to encrypt the quantum data in the earlier ``arbitrated quantum signature (AQS)'' schemes is as follows. Let an $n$-qubit quantum message $\ket{P}$ is denoted by, 
\begin{equation}\label{eq:1}
    \ket{P}=\otimes_{i=1}^{n}\ket{p_i}, \text{where} \ket{p_i}=\alpha_i \ket{0}+ \beta_i \ket{1}
\end{equation}
such that $|{\alpha_i}|^2 + |{\beta_i}|^2 = 1$ and let $K \in \{0,1\}^{2n}$ be a $2n$ bit secret key. The QOTP encryption of the message $\ket{P}$ can be represented as, 
\begin{equation}\label{eq:2}
    \ket{C}=Enc_K\ket{P}=\otimes_{i=1}^{n}X^{K_{2i}}Z^{K_{2i-1}}\ket{p_i}
\end{equation}
where $ X=\begin{pmatrix}
  0 & 1\\ 
  1 & 0
\end{pmatrix}$ and $ Z=\begin{pmatrix}
  1 & 0\\ 
  0 & -1
\end{pmatrix}$ are the Pauli operators. The QOTP decryption at the receiver's side is as follows,
\begin{equation}\label{eq:3}
\ket{P}=Dec_K\ket{C}=\otimes_{i=1}^{n}Z^{K_{2i-1}}X^{K_{2i}}\ket{c_i}
\end{equation}
where $\ket{C}=\otimes_{i=1}^{n}\ket{c_i}$. Gao \textit{\textit{et al.}} \cite{gao2011} have investigated the cryptanalysis of AQS protocols that utilize the QOTP encryption technique, particularly on efforts at forgery by the recipient, Bob and denial by the signer, Alice. Their study shows that Bob can produce numerous existential forgeries of Alice's signature in the context of a known message attack. More significantly, when the protocols are used to sign classical messages, Bob can do a universal forgery of Alice's signature. Additionally, Alice is able to effectively disown the signature she previously signed for Bob. Li and Shi \cite{Li2015CNOT} have developed a new encryption technique to address this vulnerability, which does not encrypt messages qubit by qubit as the QOTP encryption method does. They have designed the encryption system known as ``chained $CNOT$ operations encryption'', which makes use of $CNOT$ operators. The encryption algorithm is as follows,

\par Let the quantum message is denoted by $\ket{P}$ as defined in eq. (\ref{eq:1}). Let $K=(k_1,k_2,\hdots, k_n)$, which is actually a permutation of $(1,2,\hdots, n)$, be a key shared secretly between the signer and the verifier. Now, the chained $CNOT$ operation encryption is represented as
\begin{eqnarray}\label{eq:4}
\ket{C} = Enc_K\ket{P} = CNOT(p_n, p_{k_n}) CNOT(p_{n-1}, p_{k_{n-1}}) \cdots CNOT(p_1, p_{k_1})\ket{P},
\end{eqnarray}
where for the operation $CNOT(p_i, p_{k_i})$, $\ket{p_i}$ is the control qubit and $\ket{p_{k_i}}$ is the target qubit. The corresponding decryption is likewise accomplished by simply reversing the $CNOT$ operations on the encrypted message, as the operator $CNOT$ is a unitary operator. They additionally addressed a crucial aspect of their encryption technique. A single message error or an error at the $k_i$ ($i$-th position of the key) might drastically affect the encryption's outcome. This feature contributes to the security framework of AQS. 

\par In this paper, we have enhanced the encryption technique. For the encryption part, we use a controlled Unitary ($CU$) gate, which is given by
\begin{eqnarray}
CU = \begin{pmatrix}
    \mathbb{I} & O \cr
    O & U
\end{pmatrix},
\end{eqnarray}
where $\mathbb{I}$ denotes a $2\times 2$ identity matrix, $O$ a $2\times 2$ null-matrix, and $U$ a general unitary matrix given in (\ref{Ugate}).
For simplicity and to reduce the number of variables, we take $\theta = 0$ and $\phi =0$. In that case, $U$ becomes 
\begin{eqnarray}
    U(0, 0, \lambda) = \begin{pmatrix}
    1 & 0 \cr
    0 & e^{i\lambda} 
    \end{pmatrix}.
\end{eqnarray}
The Pauli-$Z$ gate can be observed as a particular instance of $U(0, 0, \lambda)$, where $\lambda = \pi$.\\
The encryption of the given message $\ket{P}$ can be explained by
\begin{eqnarray}
    \ket{C} &=& Enc_K\ket{P} \nonumber \\
     &=& CU(p_n, p_{k_n})CU(p_{n-1}, p_{k_{n-1}}) \hdots CU(p_1, p_{k_1})\ket{P},
\end{eqnarray}
where $CU(p_j, p_{k_j})$ stands for a $CU$ operator between $\ket{p_j}$ designated as the control bit, and $\ket{p_{k_j}}$ designated as the target bit. If $k_j = j$ after the permutation, then $CU(p_j, p_{k_j}) = \mathbb{I}$.

\section{Related Work}
\label{sec3}
Li \textit{et al.}  \cite{Li2009} proposed an ``arbitrated quantum signature (AQS)'' scheme using Bell states, offering improved efficiency and simplicity compared to earlier schemes based on GHZ states. By replacing GHZ states with Bell states, the scheme reduces the number of transmitted qubits and simplifies implementation, as Bell states are easier to prepare with current technology. However, its reliance on a central arbitrator introduces trust dependency, limits scalability, and creates potential vulnerabilities if the arbitrator is compromised, making it less suitable for decentralized or adversarial environments. 

Chen \textit{et al.}  \cite{Chen2017} introduced a public-key quantum digital signature scheme that integrates identity-based encryption (IBE) and a one-time pad (OTP) protocol to achieve information-theoretic security and simplify key management. The scheme employs classical bits for key pairs and quantum qubits for the signature cipher, making it efficient and compatible with existing technology. However, this protocol faces risks of forgery attacks by the PKG, and fails to resolve issues related to lost quantum signatures. To resolve the issue with respect to the PKG forgery attack, Huang \textit{et al.}  \cite{Huang2022} introduced an improved ``identity-based public-key quantum signature method''. The secret key of the signer is generated from their identity, and the protocol utilizes quantum mechanics to guarantee unforgeability, undeniability, and unconditional security. However, it still lacks a method for addressing disputes over lost quantum signatures. Here, the scheme uses $2n$ number of qubits for $n$ bit classical messages and does not include any numerical simulations, which complicates real-world validation. To address these limitations, Prajapat \textit{et al.}  \cite{Sunil2024} presented an identity-based ``quantum-designated verifier signature (QDVS)'' scheme that incorporates resistance to eavesdropping, ensures non-repudiation, allows designated verification, and safeguards against source-hiding attacks. The scheme employs an entangled state for signing and verification processes without necessitating the verifier to compare quantum states. In addition, it includes a comprehensive security comparison in relation to existing schemes and offers simulation code along with measurement information for various phases, including the processes of signature generation and verification. However, the protocol still faces vulnerabilities to Pauli-operator attacks and relies on numerical simulations that exclusively use computational basis states, thus not fully leveraging the advantages of quantum superposition. 

Mohanty \textit{et al.} \cite{mohanty2024} introduced an ``identity-based signature (IBS)'' presented a quantum cryptography-based identity-based signature (IBS) scheme aimed at thwarting Pauli-Operator attacks by utilizing a generalized form of the unitary operator, in contrast to standard quantum operators. The implementation of the scheme was carried out using the IBM Qiskit quantum simulator, with accompanying code available in Qiskit and Jupyter Notebook. Although their paper offers a comprehensive performance evaluation relative to existing schemes, the suggested use case for secure email communication does not have validation from a quantum point of view.

Zhang \textit{et al.}  \cite{Zhangimproved} introduced a novel quantum encryption method called key-controlled chained $CNOT$ encryption and used it to develop an improved AQS protocol. Their scheme is shown to resist forgery and disavowal attacks while requiring only a single state. Additionally, it includes a security comparison with other schemes. However, the protocol does not include a numerical simulation in the quantum scenario.

\section{Proposed Quantum Signature Scheme} \label{sec4}
This section introduces a new quantum signature scheme. In this scheme, we consider $m$ parties, labelled as Alice$_1$, Alice$_2$,$ \hdots $, Alice$_m$, who serve as the signers; Bob, who functions as the signature verifier; and a trusted third party, Key Generation Center (KGC), who not only generates the secret key for each Alice$_i$ (\(1 \leq i \leq m\)) but also assist Bob in verifying the signatures. Our proposed scheme is organized into four different phases: 1) ``initialization phase'', 2) ``key generation phase'', 3) ``signing phase'' and 4) ``verification phase''. The following describes each of these stages in order.

\subsection{Initialization phase}
In this proposed scheme, the KGC (Key Generation Center) is treated as a fully trusted authority. It is assumed that the KGC is under a physical locking mechanism as in \cite{Bertino4384503,Wazid8070995}, which is realistic in the real-world scenario. This is significant, because  regardless of the cryptographic security of the KGC, an individual with physical access to the machine could potentially steal private keys or set up hardware backdoors. In this way, the risk of vulnerabilities in the KGC is significantly reduced. 

This phase involves the following steps:
\begin{itemize}
    \item KGC will choose a uniformly distributed one-way hash function \( H: \{0,1\}^* \rightarrow \{0,1\}^n \) and publicly broadcast it to all parties participating in the protocol.
    \item Alice$_i$ prepares two copies of the $n$-qubit quantum message $\ket{P^i}=\otimes_{j=1}^{n}\ket{p^i_j}$ that is to be signed, where each qubit $\ket{p^i_j}=\alpha^i_j \ket{0}+ \beta^i_j \ket{1}$ such that $|{\alpha^i_j}|^2 + |{\beta^i_j}|^2 = 1$.
\end{itemize}

\subsection{Key generation phase}
In this phase, KGC generates and shares the secret keys $K_1,K_2,\hdots K_m$ (where each $K_i \in \{0,1\}^n$ for $1 \leq i \leq m$) to Alice$_1$, Alice$_2$,$ \hdots $, Alice$_m$ respectively and $K_B$ (where $K_B \in \{0,1\}^n$) to Bob using ``quantum key distribution (QKD)'' protocol. 
\begin{itemize}
    \item After receiving the secret keys, the signers Alice$_1$, Alice$_2$, $\hdots$, Alice$_m$ compute $K'_1,K'_2,\hdots K'_m$ respectively which they keep secret as well. Each of $K'_1,K'_2,\hdots K'_m$ is a permutation of $(1,2,\hdots, n)$, computed as follows. Based on the positions of 0 and 1 in the key $K_i$, Alice$_i$ retrieves the key $K'_i$. For $1 \leq i \leq m$, let the key $K'_i$ be denoted as ($k'_{i1}, k'_{i2}, \hdots, k'_{in}$) where $k'_{i1}$ is the position of the first 0 in the key $K_i$, $k'_{i2}$ is the position of the second 0 in the key $K_i$, in this way $k'_{ij}$ is the position of the last 0 in the key $K_i$, when $j$ is the total number of 0 in $K_i$. Similarly, $k'_{i(j+1)}$ is the position of the first 1 in the key $K_i$, and $k'_{in}$ is the position of the last 1 in the key $K_i$. 
    \item Likewise, Bob computes the key $K'_B$ from the key $K_B$. 
\end{itemize}
\begin{example} \label{Ex1}
If the shared secret key $K_i$ is 11010, $K'_i$ will be (3, 5, 1, 2, 4). 
\end{example}

\subsection{Signing phase}
The signing phase consists of two sub-phases: a) encryption phase and b) signature generation phase. These are discussed in detail below.
\subsubsection{Encryption phase}
\begin{itemize}
    \item Alice$_i$ chooses $n$ values of $\lambda_i \in [0, \pi]$ (Let's say $\lambda^1_i,\lambda^2_i, \hdots, \lambda^n_i$) uniformly randomly and shares with KGC using quantum authentication protocol  \cite{xin2015quantum}.
    \item Next, in order to generate the signature of the message $\ket{P^i}$, Alice$_i$ first encrypts one copy of $\ket{P^i}$ as follows:
\begin{eqnarray}
\label{eq:9}
    \ket{C^i} &=& CU(0,0,\lambda^n_i)(p_n, p_{k'_{in}})\hdots CU(0,0,\lambda^{n-1}_i)(p_{n-1}, p_{k'_{i(n-1)}}) \hdots {}\nonumber\\ &&
CU(0,0,\lambda^1_i)(p_1, p_{k'_{i1}})\ket{P^i}
\end{eqnarray}
\end{itemize}

\subsubsection{Signature generation phase}
\begin{itemize}
\item Then the signer Alice$_i$ generates the signature $\ket{S^i}$ as follows
\begin{equation}
\label{eq:10}
    \ket{S^i} = \otimes_{j=1}^n U(0,0,\lambda^j_i) \ket{C^i_j}
\end{equation}
\item Next, Alice$_i$ computes $h_i=H(K_i)$ and transmits the sequence $\{\ket{P^i}, \ket{S^i}, h_i\}$ to Bob.
\end{itemize}

\begin{figure*}[htb!]
    \centering
    \includegraphics[width=0.9\linewidth]{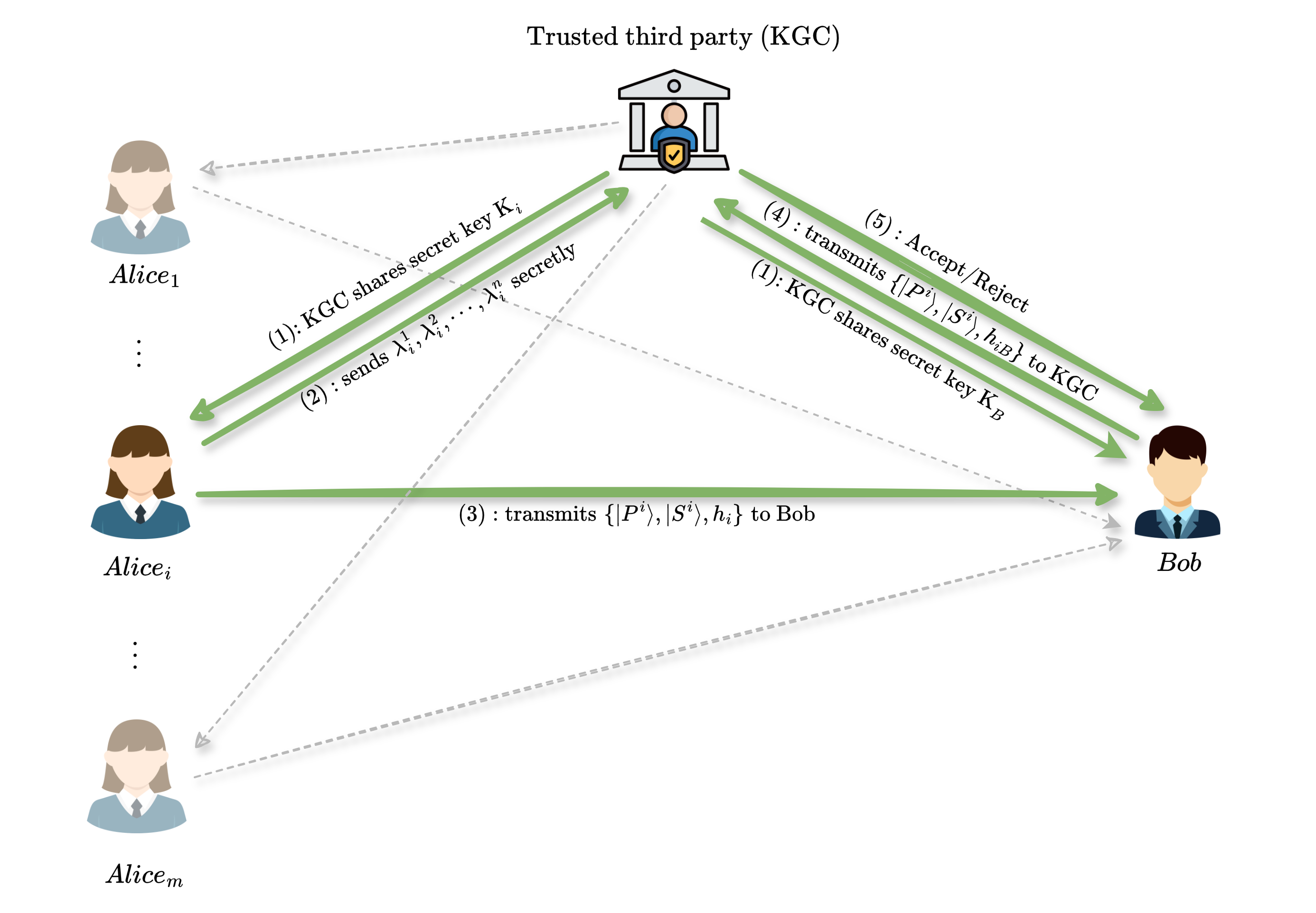}
    \caption{Schematic diagram of the proposed quantum signature scheme.}
    \label{fig:scheme diagram}
\end{figure*}

\subsection{Verifying phase} 
Once Bob receives the sequence $\{\ket{P^i}, \ket{S^i}, h_i\}$ from Alice$_i$, Bob reach out to KGC to verify the authenticity of the received message-signature pair, as outlined in the following steps: 
\begin{itemize}
    \item Bob computes $h_{iB}=H(h_i \oplus K_B)$ and transmits $\{\ket{P^i}, \ket{S^i}, h_{iB}\}$ to KGC.
    \item KGC first computes $h^*_i = H(H(K_i) \oplus K_B)$, if $h^*_i \neq h_{iB}$, he rejects the signature $\ket{S^i}$; otherwise, he decrypts $\ket{S^i}$ to $\ket{C^i}$ using the knowledge of $\lambda^j_i$'s for $1 \leq j \leq n$. Then from $\ket{C^i}$ he gets back $\ket{P^i}$ using the secret key $K'_i$ and $\lambda^j_i$ as follows:
    \begin{eqnarray}
    \ket{P^i} &=& CU^\dag(0,0,\lambda^1_i)(p_1, p_{k'_{i1}}) CU^\dag(0,0,\lambda^{2}_i)(p_{2}, p_{k'_{i2}})\hdots {}\nonumber\\&&  CU^\dag(0,0,\lambda^n_i)(p_n, p_{k'_{in}})\ket{C^i}
\end{eqnarray}
\item KGC then verifies whether the decrypted $\ket{P^i}$ matches the original message $\ket{P^i}$ as forwarded by Bob, utilizing the quantum swap test algorithm  \cite{Buhrman2001}. If they match, KGC responds with ``Yes" to Bob, and Bob accepts the signature as valid; otherwise, KGC responds with ``No," and Bob rejects the signature.
\end{itemize}
For each party Alice$_i$, once verification has been completed successfully, KGC stores \{($\lambda^1_i,\lambda^2_i, \hdots, \lambda^n_i$), $h_{iB}$\} as signature proof, which can be used to settle disputes involving lost signatures. The streamlined steps of our proposed signature scheme are depicted in Fig. \ref{fig:scheme diagram}. In the following section, the security attributes of this proposed scheme are examined.

\section{Security analysis and discussion} \label{sec5}
Two security conditions must be satisfied by a secure quantum signature scheme: non-forgery and non-repudiation. In this section,we demonstrate that our scheme upholds these properties and remains robust against various other potential attacks.

\subsection{Secrecy of the private key}
\label{sec5.1}
First, we assess the confidentiality of the secret key in this protocol. The KGC acts as a reliable third party and has no incentive to reveal the signer Alice$_i$'s secret key. Additionally, we will demonstrate that, in our protocol, even the verifier cannot deduce the signer's secret key from the information provided to him through the public channel. Consequently, it becomes impossible for any external adversary to compromise the signer's private key.
\par In the key generation phase, KGC generates and shares the secret key $K_i$  to the signer Alice\(_i\) (where each \( K_i \in \{0,1\}^n \) for \( 1 \leq i \leq m \)) and \( K_B \) (where \( K_B \in \{0,1\}^n \)) to the verifier Bob using the QKD protocol. Thus, in this process, the principles of quantum mechanics guarantee that it is impossible for an eavesdropper to intercept and successfully extract the private keys. Next, Alice$_i$ computes $K'_i$ from the key $K_i$ and Bob computes $K'_B$ from the key $K_B$. Since the keys $K_i$ and $K_B$ are secret, the keys $K'_i$ and $K'_B$ remain secret as well. 
\par Next, during the signing phase, the verifier, Bob, may try to decrypt the secret key from the signature he receives. Nevertheless, this approach is also ineffective, as Bob lacks any information about the secret data $\{\lambda^1_i,\lambda^2_i, \hdots, \lambda^n_i\}$ and must also send the signature $\ket{S^i}$ to KGC for verification purposes.
\par Finally, any malicious adversary might attempt to decrypt the signer's secret key $K_i$ using the knowledge of $h_i$. However, since the output of the hash function $H$ is uniformly distributed over the space $\{0,1\}^n$, the attacker is left with only a negligible probability $\frac{1}{2^n}$ of guessing the secret key.

\subsection{Non-forgery}
\label{sec5.2}
A legitimate quantum signature scheme must guarantee that only the designated signer, such as Alice$_i$, can produce their signature, ensuring that no other party can create the signature on their behalf. Consider Eve to be an external adversary attempting to counterfeit Alice$_i$'s signature. Eve may attempt to do so using one of the following two methods.
\par Firstly, Eve might attempt to forge a different signature $\{\ket{P^i}, \ket{S_*^i}\}$ after intercepting a valid quantum signature $\{\ket{P^i}, \ket{S^i}\}$. In other words, given the valid signature $\{\ket{P^i}, \ket{S^i}\}$, Eve tries to impersonate Alice$_i$ by forging a new signature $\{\ket{P^i}, \ket{S_*^i}\}$ for the same message $\ket{P^i}$. However, Eve cannot successfully forge the signature because generating a legitimate signature requires the secret information of $K_i$ and \{$\lambda^1_i, \lambda^2_i, \dots, \lambda^n_i$\}, which are only accessible to Alice$_i$ and KGC. Therefore, the KGC can detect any forged signature $\ket{S_*^i}\}$  at the time of verification phase, as decrypting $\ket{S_*^i}\}$ will produce $\ket{P_*^i}$, which will not match the original message $\ket{P^i}$. Thus, Eve is unable to commit this type of forgery. 
\par In the second forgery scenario, Eve chooses a message $\ket{P_*^i}$ and executes a cryptographic operation on it to obtain a signature $\ket{S_*^i}$. However, a valid signature must satisfy the Eqs.  \ref{eq:9} and  \ref{eq:10} which requires the knowledge of $K_i$ and \{$\lambda^1_i, \lambda^2_i, \dots, \lambda^n_i$\}. Thus, by a similar analysis to that above, we can conclude that this type of forgery is also impossible for Eve, and consequently, our scheme is secure against forgery attacks.

\subsection{Non-repudiation}
\label{sec5.3}
An essential aspect of a signature scheme is non-repudiation, also called resistance to repudiation. A secure signature protocol must ensure that the signer cannot withdraw or deny their previously signed valid signature.
\par Since KGC is a trusted third party, it is assumed that the KGC will never disclose the private key or other confidential information of any user, such as Alice$_i$. Alice$_i$ uses her private key to generate the corresponding quantum signature to sign a message. The verifier, Bob, validates the quantum signature with the assistance of the KGC, leveraging the secret information specific to each signer, Alice$_i$. As established in Section  \ref{sec5.2}, the proposed scheme is resistant to forgery attacks. Therefore, once the quantum signature successfully passes the verification phase, Alice$_i$ cannot repudiate it, as she is the only one capable of generating the signature using her private key.
\par Moreover, the majority of the quantum signature protocols that are available in the literature cannot resolve disputes regarding lost quantum signatures. Typically, once the signature gets verified in a quantum signature protocol, the signer and the verifier lose the signature because the quantum state changes. In such cases, the signer might deny ever generating the signature, and the verifier might deny having verified it. This kind of disagreement can be arbitrated by our quantum signature scheme because for each party Alice$_i$, once verification has been completed successfully, KGC stores \{($\lambda^1_i,\lambda^2_i, \hdots, \lambda^n_i$), $h_{iB}$\} as signature proof. In our scheme, even though the verification is conducted with the assistance of the KGC, Bob cannot deny having received the signature from Alice$_i$. This is because KGC stores the information $h_{iB}$, which includes the information of Bob's secret key that he provided to KGC after receiving the signature from Alice$_i$. As a result, Alice$_i$ is unable to argue that she created the signature on $\ket{P^i}$. Similarly, the verifier, Bob, is also unable to deny ever having verified the quantum signature.

\subsection{Resistance to Pauli operator forgery attacks}
The proposed protocol is intrinsically resistant to Pauli operator forgery attacks, such as the one proposed by Gao \textit{et al.} \cite{gao2011}, which target QOTP-based AQS protocols. These attacks exploit the commutation relations between Pauli operators and QOTP encryption, particularly the ability to apply the same Pauli operation to both the message and signature to create an undetectable forged pair. In contrast, our scheme does not rely on QOTP or any qubit-wise symmetric encryption. Instead, we adopt a chained controlled-unitary ($CU$) gate-based encryption and the private parameters \{$\lambda^1_i, \lambda^2_i, \dots, \lambda^n_i$\} render the effect of any Pauli operation highly nonlinear and dependent on both the structure and ordering of $CU$ gates. Since Pauli operators do not commute predictably with arbitrary unitary operators, especially within a chained $CU$ configuration, any unauthorized Pauli transformation on the signature would propagate unpredictably through the $CU$ layers and result in a verifiable mismatch during the KGC’s decryption and quantum state comparison process. Therefore, unlike QOTP-based schemes, our encryption inherently thwarts Pauli operator forgery by invalidating the attacker’s ability to construct existential or universal forgeries without detection.

\subsection{Impersonation attack resistance}
Due to the integration of quantum cryptographic primitives such as quantum key distribution (QKD) and quantum authentication, impersonation attacks are effectively mitigated in this signature scheme. Each signer, Alice$_i$, receives a distinct secret key $K_i$ from the Key Generation Centre (KGC) via QKD, which guarantees unconditional security and ensures that no adversary, including an impersonator, can intercept or replicate the key without detection. Additionally, each signature incorporates a unique, randomly chosen set of parameters $(\lambda^1_i, \lambda^2_i, \dots, \lambda^n_i)$ that are transmitted to the KGC using a quantum authentication protocol, preventing any adversary from forging valid signature components without knowledge of both the key and these parameters. Since the signature verification also involves a secure comparison of quantum states using the swap test and hash verification through $h_{iB} = H(H(K_i) \oplus K_B)$, an attacker cannot impersonate a valid signer without being detected. Thus, any attempt to forge a message or signature by impersonating a signer will fail due to the fundamental principles of quantum mechanics and the cryptographic structure of the protocol.

\subsection{Man-in-the-Middle (MITM) attack resistance}

The proposed quantum signature scheme is secure against Man-in-the-Middle (MITM) attacks due to several quantum and classical cryptographic safeguards. First, all secret keys between the signers (Alice$_i$), the verifier (Bob), and the trusted third party (KGC) are distributed using quantum key distribution (QKD), which inherently detects eavesdropping due to the disturbance introduced in quantum states upon measurement. Second, the quantum authentication protocol used to transmit the $\lambda_i^j$ parameters from the signers to KGC further ensures message integrity and origin authentication, preventing impersonation or tampering by an adversary. Third, the verification process relies on hash comparisons involving secret keys and the use of a quantum swap test for validating the integrity of the quantum message, ensuring any attempt to forge or alter messages is detected. Finally, because quantum information cannot be copied (no-cloning theorem) or measured without disturbing it, an adversary in the middle would be unable to intercept and retransmit valid quantum states without detection. Collectively, these features make MITM attacks practically impossible in the proposed scheme.

\section{Implementation on an Actual Back-end Device}
\label{sec6}
We execute our proposed protocol on a quantum simulator that mimics an IBM Quantum back-end device. Then, we implement and run it on an actual back end.

\subsection{Description of the System}
\begin{figure*}[htb!]
    \centering
    \includegraphics[width=\linewidth]{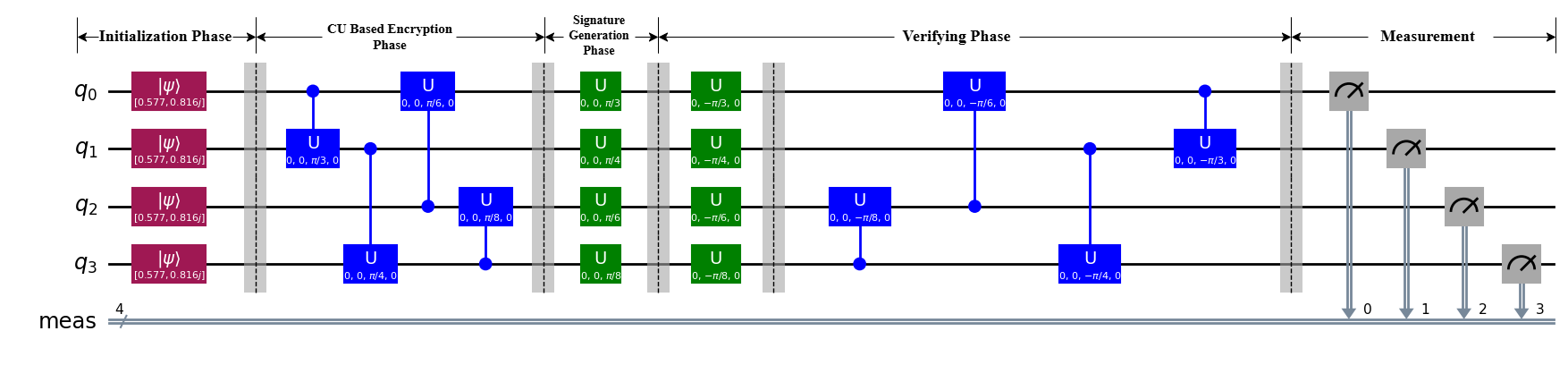}
    \caption{The final circuit of the given protocol in the mentioned scenario. Here, the brown colour elements depict the initialization acting on the initial qubits. The initialized message is $\ket{P^1} = \left( \frac{1}{\sqrt{3}} \ket{0}+ i\sqrt{\frac
    {2}{3}}\ket{1}\right)^{\otimes 4}$. The blue colour elements suggest the control-unitary gates, and the green colour elements are the unitary operators acting on the corresponding qubits. The measurements are done at the end of the circuit.}
    \label{fig:ckt}
\end{figure*}

We consider the most trivial case with just a single Alice, Bob and the KGC. In the key generation phase, KGC generates the secret key $K_1 = 1010$ and shares it with Alice\textsubscript{1}. Therefore, $K_1' = (1,3,0,2)$. Now, let us take a four-qubit message as
\begin{eqnarray}
    \ket{P^1} = \left( \frac{1}{\sqrt{3}} \ket{0}+ i\sqrt{\frac
    {2}{3}}\ket{1}\right)^{\otimes 4}
\end{eqnarray}
which needs to be signed by Alice\textsubscript{1}. Alice\textsubscript{1} randomly chooses $\lambda_1$ as $(\frac{\pi}{3}, \frac{\pi}{4}, \frac{\pi}{6}, \frac{\pi}{8})$. Therefore, the encryption looks like
\begin{eqnarray}
    \ket{C^1} = CU(0,0,\frac{\pi}{8})(3,2)CU(0,0,\frac{\pi}{6})(2,0) CU(0,0,\frac{\pi}{4})(1,3) CU(0,0,\frac{\pi}{3})(0,1)\ket{P^1}. \nonumber
\end{eqnarray}
This leads to the following signature generated by Alice\textsubscript{1}
\begin{eqnarray}
    \ket{S^1} =  \otimes_{j=0}^3 U(0,0,\lambda^j_i) \ket{C^1_j}. 
\end{eqnarray}
Alice\textsubscript{1} sends $\{\ket{P^1}, \ket{S^1} , h_1\}$ to Bob and $\ket{P^1}$ to KGC. On the other hand, Bob sends $\{\ket{S^i}, h_{1B} \}$ to KGC. We consider that there was no security breach during this process. Therefore, KGC decrypts $\ket{C^1}$ and then $\ket{P^1}$. The final circuit is shown in Fig.  \ref{fig:ckt}. We executed the measurement two different times and compared the results.

\subsection{Implementing on a Quantum Simulator}
For this case, we have used the open-source software Qiskit AerSimulator (\textit{Qiskit Aer 0.15.0}). At first, we measure the circuit after preparing the message $\ket{P^1}$ using this simulator. This is depicted in Fig.  \ref{fig:comparison} in yellow bars. Now, after the protocol runs, we take another measurement, and the values are depicted in the blue bars of the figure. Table  \ref{tab:1} mentions the software and hardware specifications.
\begin{table}
\caption{Specifications for IBM Qiskit quantum simulation}
    \label{tab:1}
    \centering
    \begin{tabular}{|c|c|}
    \hline
     Qiskit Version  & 1.3.1\\
     \hline
     Qiskit Element    & Qiskit Aer (version 0.15.1)\\
     \hline
     Simulator    & QASM\\
     \hline
     Python version   & 3.13.1\\
     \hline
     Local OS   & Fedora Linux 41\\
     \hline
     Local Hardware  & Intel\textsuperscript{\textregistered} Core\textsuperscript{TM} i5, RAM 8.00 GB\\
     \hline
    \end{tabular}
    
\end{table}
\begin{table}
\caption{Hardware specifications on a real quantum computer.}
    \label{tab:2}
    \centering
    \begin{tabular}{|c|c|}
    \hline
     Qiskit Version  & 1.3.1\\
     \hline
     Quantum Processing Unit  & ibm\_brisbane\\
     \hline
     No. of Qubits    & 127\\
     \hline
     Processor Type   & Eagle r3\\
     \hline
     CLOPS (Circuit Layer Operations per Second)   & 180K\\
     \hline
     1Q error  & 1.2459e-2\\
     \hline
     2Q error  & 3.68e-3 (best), 1.76e-2 (layered)\\
     \hline
     Median ECR Error & 7.086e-3\\
     \hline
    \end{tabular}
    
\end{table}

\subsection{Implementing on a Real Back-end Device}
\begin{figure*}[htb!]
    \centering
    \includegraphics[width=0.79\linewidth]{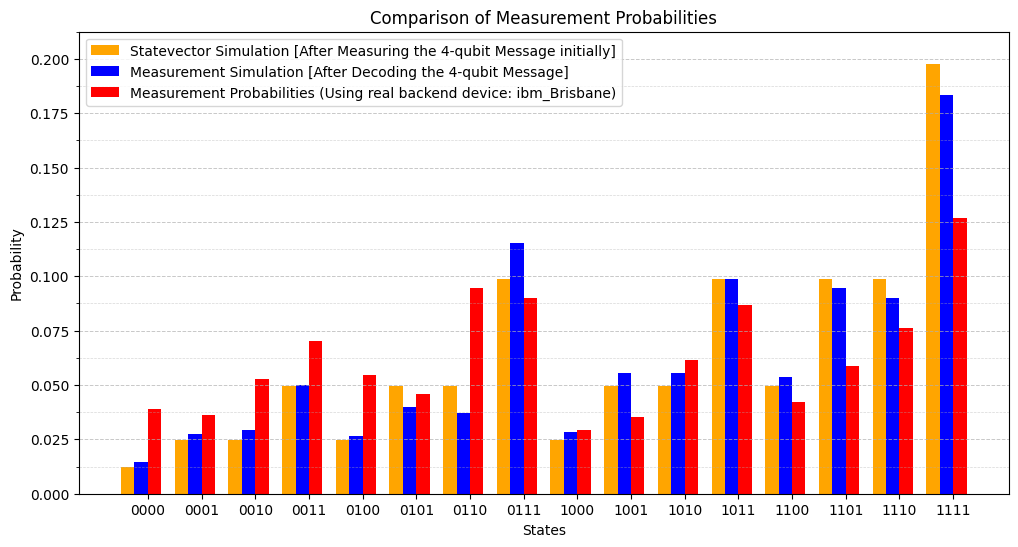}
    \caption{The histogram for comparison of the measurement probabilities. (a) In the first bin (in yellow), the message is measured after preparing the 4-qubit message. (b) The next bin (in blue) depicts the states' probability after the whole protocol using the Quantum AerSimulator. (c) The final bin (in red) indicates the states' probability (by running 1024 times) using the real backend device \textit{ibm\_brisbane}.}
    \label{fig:comparison}
\end{figure*}

The proposed scheme is implemented on a real backend device \textit{ibm\_brisbane} provided by IBM Quantum Experience. This protocol was executed 1024 times on this actual backend device. Table \ref{tab:2} mentions the hardware details of \textit{ibm\_brisbane}. The results of the probability of the states after the measurement are shown in Fig. \ref{fig:comparison} in red bars. This measurement clearly shows the distinction, while we want to retrieve the message after the whole protocol. For example, we have taken the initial state as $\ket{P^1} = \left( \frac{1}{\sqrt{3}} \ket{0}+ i\sqrt{\frac{2}{3}}\ket{1}\right)^{\otimes 4}$. For this state, the amplitude of state $\ket{0110}$ is -0.22222 and hence the measurement probability is 0.04938, which is shown as the yellow bar. Now, when we simulate the same circuit in the Aer Simulator, the measurement probability of the same state is 0.037, and it is shown as the blue bar. Meanwhile, in the case of a real backend device, the measurement probability is 0.09473, and it is shown as the red bar. This affirms the deviation between the initial measurement and the final measurement after retrieving the message.

\subsection{Efficiency and scalability}
To evaluate the efficiency and scalability of our protocol, we analyzed the gate complexity and circuit depth involved in the signing and verification phases. For a four-qubit quantum message, our scheme utilizes four controlled-unitary ($CU$) gates and four parameterized single-qubit unitary ($U$) gates for encryption and a corresponding set for decryption. Additionally, the circuit includes initialization and measurement stages. The high-level quantum gate count is summarized in Table \ref{tab:gatecount}.

\begin{table}[h!]
\centering
\caption{High-level gate count per 4-qubit message}
\label{tab:gatecount}
\begin{tabular}{@{}lc@{}}
\toprule
\textbf{Gate type}            & \textbf{Count} \\ \midrule
Initialization (Initialize)   & 4              \\
$CU$ Gates (Encryption)         & 4              \\
$U$ Gates (Signature creation)  & 4              \\
$U^\dagger$ Gates (Decoding)  & 4              \\
$CU^\dagger$ Gates (Decryption) & 4            \\
Measurement                   & 4              \\ \midrule
\textbf{Total}                & \textbf{24}    \\ \bottomrule
\end{tabular}
\end{table}

Each $CU$ gate is a non-Clifford operation and requires decomposition into several basic gates such as $CNOT$ and single-qubit rotations, making it computationally more expensive compared to Clifford-only gates. While each `initialize' gate is abstracted as a single operation in high-level quantum circuit design, its implementation also contributes to the depth when transpiled. The approximate circuit depth accounting for non-parallelizable dependencies is provided in Table~\ref{tab:depth}.

\begin{table}[h!]
\centering
\caption{Approximate circuit depth by phase}
\label{tab:depth}
\begin{tabular}{@{}lc@{}}
\toprule
\textbf{Circuit phase}        & \textbf{Depth (Layers)} \\ \midrule
Initialization               & 1                       \\
$CU$ Encryption                & $\sim$4                 \\
$U$ Gate application           & 1                       \\
$U^\dagger$ Gate application & 1                       \\
$CU$ Decryption                & $\sim$4                 \\
Measurement                  & 1                       \\ \midrule
\textbf{Total depth}         & \textbf{$\sim$12}       \\ \bottomrule
\end{tabular}
\end{table}

Compared to the chained $CNOT$-based encryption scheme (KCCC) in Zhang \textit{et al.}'s \cite{Zhangimproved} protocol, which achieves linear gate complexity and shallow circuit depth using only Clifford gates ($CNOT$ and Hadamard), our $CU$-based design introduces additional depth and gate overhead. Specifically, each $CU$ gate incurs a decomposition depth $d > 1$, and for an $n$-qubit message, the total depth becomes $\mathcal{O}(d \cdot n)$ compared to $\mathcal{O}(n)$ in KCCC. However, this increased depth enables stronger entanglement structure and more secure signature binding. Despite the overhead, the $CU$-based scheme maintains scalability for moderate $n$, as all operations remain polynomial in $n$. This trade-off favours enhanced security at the cost of computational complexity.

\subsection{Analysis of Obtained Result and Adaptation of the Protocol in NISQ Era} In the case of the quantum processing unit (QPU), there is 1Q error. A 1Q error, or 1-qubit gate error, is the likelihood that a single-qubit quantum gate (such as $X, Y, Z, H, S, T,$ etc.) may generate an inaccurate output because of flaws in the control procedures or quantum hardware. There are also 2Q (two-qubit gate errors) errors, which indicate that an error occurs when a quantum computer performs a gate involving two qubits. In our designed protocol, controlled unitary gates are present. This leads to the 2Q (two-qubit) error. Furthermore, the median ECR error reflects the average gate fidelity, which is the average similarity between the ideal and actual output states of a gate, taken over all possible input states. It is clear that these errors constrain the retrieval of the desired message with 100\% fidelity.\\ 
In the case of modern NISQ (Noisy Intermediate-Scale Quantum) devices, these errors lead to massive disputes when the circuit contains more than 1000 gates \cite{Preskill2018}. Hence, quantum error correction technology such as Quantum Error Correction Codes, Dynamical Decoupling and Echo techniques, etc., is necessary to recover the original message. In the case of IBM hardware, dynamical decoupling and echo techniques are used to handle 2Q and ECR errors, which have low implementation complexity compared to other techniques. Echoed cross-resonance (ECR) is a method for cancelling undesirable noise during two-qubit gates, while dynamical decoupling uses precisely timed pulses to reduce some of the noise in qubit interactions. IBM devices use this technique to reduce the noise in the hardware \cite{Niu2022}. However, this imposes a heavy cost on a number of qubits as well as a number of gates. Hence, it is important to research further about the noise-resilient algorithm of the proposed scheme.

\section{Comparative Study}
\label{sec7}
In Table  \ref{comparison table}, we highlight the efficiency advantages of our proposed quantum signature scheme compared to other quantum signature schemes based on the selected key attributes: message space, encryption depth, resistance to forgery using Pauli operators, key reusability, and the ability to arbitrate the loss of quantum signatures. 

The scheme by Gottesman and Chuang \cite{gottesman2001quantum} supports only classical message space and uses a quantum one-way function based on nearly orthogonal states; however, it neither resists Pauli operator forgery nor allows key reusability or arbitration for lost signatures. Chen \textit{et al.}'s scheme \cite{Chen2017} also targets classical messages and introduces a one-time pad with digest mapping but still lacks forgery resistance and arbitration capability, although it improves key reusability. Huang \textit{et al.} \cite{Huang2022} and Xin \textit{et al.} \cite{Xin2020} enhance forgery resistance using Clifford operations and entanglement-based mechanisms respectively, and both support key reusability; however, only Huang's and Xin’s 2020 schemes provide a way to arbitrate lost signatures. Xin \textit{et al.}'s earlier work in 2019 \cite{Xin2019} also resists forgery and supports key reuse but lacks arbitration. Li \textit{et al.} \cite{Li2009} expands support to both quantum and classical messages using QOTP, though it does not resist forgery or enable arbitration. Zhang \textit{et al.}'s \cite{Zhangimproved} improved scheme supports both message types and offers forgery resistance and key reusability via chained $CNOT$ and Hadamard gates but still lacks arbitration. In contrast, our proposed scheme provides a holistic solution by supporting both quantum and classical messages, applying deep encryption via chained controlled-unitary operations, ensuring resistance to Pauli-based forgery, enabling key reusability, and uniquely facilitating arbitration for lost quantum signatures, making it the most comprehensive among existing approaches.

\begin{table*}[!htb]
\caption{Comparative analysis} \label{comparison table} 
\centering
\resizebox{\textwidth}{!}{%
\begin{tabular}{|l c p{6cm} c c c| } \hline
%\rowcolor{red!9}
Scheme & Message space & Encryption Depth & Ability to resist  & Reusability  & Ability to  \\ 
       &               &                  &  forgery with       &   of keys   & arbitrate loss of \\ 
       &               &                  &  Pauli operators    &             & quantum signature \\ \hline \hline
\rowcolor{gray!9}
 \cite{gottesman2001quantum} & C & Quantum one-way function using nearly orthogonal quantum states & No & No & No\\ \hline
 \cite{Chen2017}  & C & One-time pad encryption + QOWF mapping to digest space & No & Yes & No \\ \hline 
\rowcolor{gray!9}
 \cite{Huang2022} & C & Key-controlled Clifford operators (`I'-QOTP) & Yes & Yes & Yes\\ \hline 
\cite{Xin2019}& C  & QOWF + identity-based hash-digest with OTP & Yes & Yes & No \\ \hline
\rowcolor{gray!9}
\cite{Xin2020}& C & Bell-state encoding with identity-based hash and entanglement checking & Yes & Yes & Yes \\ \hline
\cite{Li2009}& Q, C & QOTP & No & Yes & No \\ \hline
\rowcolor{gray!9}
\cite{Zhangimproved}& Q, C & Key-controlled chained $CNOT$ + Hadamard gates & Yes & Yes & No \\ \hline
Proposed & Q, C & Chained controlled-unitary operations leveraging phase and amplitude-level encryption & Yes  & Yes & Yes\\ \hline
\end{tabular}}
\end{table*}

\section{Conclusion}
\label{sec8}
In this work, we introduced the chained controlled unitary operations encryption, which was designed to encrypt quantum message ensembles. Unlike QOTP, which encrypts messages qubit by qubit, this method is better suited for AQS. The security of our proposed scheme is ensured by the chained controlled unitary operations encryption. Security analysis demonstrates that the protocol is immune to existing forgery and disavowal attacks. Lastly, we implement the proposed scheme on a real backend device provided by IBM Quantum Experience. 

The limitation of the proposed scheme is that it's security excessively depends on fully trusted KGC, which may lead to create centralized vulnerability points. In our scheme, the KGC is under physical locking system and can not be compromised easily. However, in future, we would like to extend the proposed scheme by removing much dependency on the centralized KGC to enhance the security in the quantum signature protocol. Additionally, we anticipate that our study will be useful in the future for designing secure AQS protocols.

\bmhead{Acknowledgment} 
We acknowledge the use of ``IBM Qiskit framework for this work. The views expressed are those of the authors and do not reflect the official policy or position of IBM or the IBM Quantum team". S.R. would like to acknowledge DST-India for the INSPIRE Fellowship (IF220695) support.

\bibliography{sn-bibliography}% common bib file


\end{document}